\documentclass[conference]{IEEEtran}
\IEEEoverridecommandlockouts
\usepackage{cite}
\usepackage{amsmath,amssymb,amsfonts}
\usepackage{algorithmic}
\usepackage{graphicx}
\usepackage{textcomp}
\usepackage{xcolor}
\usepackage{setspace}
\def\BibTeX{{\rm B\kern-.05em{\sc i\kern-.025em b}\kern-.08em
		T\kern-.1667em\lower.7ex\hbox{E}\kern-.125emX}}
\usepackage{subfigure}
\newtheorem{my_theorem}{Theorem}

\newtheorem{my_lemma}{Lemma}

\usepackage[margin=0.707in]{geometry}
\title{Performance of Dual-Hop Relaying for THz-RF Wireless Link}
\author{
	\IEEEauthorblockN{Pranay Bhardwaj and S. M. Zafaruddin}\\
	\IEEEauthorblockA{ Deptt. of Electrical and Electronics Engineering, 
		BITS Pilani, Pilani-333031, Rajasthan, India.\\ Email: \{pranay.bhardwaj, syed.zafaruddin\}@pilani.bits-pilani.ac.in}
	
	\thanks{  }
}

\begin{document}
	\maketitle 
	\begin{abstract}
 The use of Terahertz (THz) frequency bands for data transmissions between the core network and an access point can be promising for next generation wireless systems. In this paper, we analyze the performance of a dual-hop relaying for THz-RF wireless link for backhaul applications. Considering the $\alpha-\mu$ fading channel and a statistical model of pointing errors, we derive a novel closed-form expression of the cumulative distribution function (CDF) of the signal-to-noise ratio (SNR) for the THz link, which is also valid for non-integer values of $\mu$. Using the CDF, we derive analytical expressions of the end-to-end SNR and lower bound on ergodic capacity of a decode-and-forward (DF) assisted THz-RF relaying in terms of system parameters. Using analytical results of  the direct link and  computer simulations, we demonstrate that the THz-RF relaying is a viable technology for wireless backhaul, providing  a significant increase of  almost $25 \%$ in the spectral efficiency, compared to the direct transmissions.
	\end{abstract}		
	\begin{IEEEkeywords}
		 Decode and forward, Ergodic rate, Outage Probability, Pointing error, Relaying, Signal to noise ratio, Terahertz.
	\end{IEEEkeywords}	
\section{Introduction}
Network densification is a potential technology to support the widespread proliferation of high data rate applications for a large number of devices in a network. This densification can be realized by adding more cell sites, including radio access networks (RAN), macro sites, small cell deployments, and the cell-free architecture. The devices are usually connected over radio frequency (RF) to a nearby access point (AP), which transports data to the core network through a high-speed back-haul link. The wire-line back-haul consists of digital subscriber lines (DSL) or optical fiber, which may not be available in some adverse situations. In contrast to the wireless back-haul link over the RF frequencies, Terahertz (THz) wireless systems can be a promising alternative for high data rate transmission for a large number of users. Thus, the performance analysis of the THz-RF relaying scheme is highly desirable.

The THz spectrum provides tremendously high, unlicensed bandwidth which can be useful for next generation of wireless technologies \cite{Koenig_2013_nature,Elayan_2019}. However, it suffers from pointing errors when there is a misalignment between transmitter and receiver antenna beams at higher frequencies in addition to the signal fading and higher path loss. The path loss in the THz band is higher due to molecular absorption of the signal with extremely small wavelength \cite{Kokkoniemi_2018}. Recently, there have been an increased research interest to model the effect of channel fading and  pointing errors on the THz transmission \cite{Boulogeorgos_Analytical,Priebe_2011,Jornet_2011,Wu_2019,Cheng_2020,Olutayo_2020}. The authors in \cite{Boulogeorgos_Analytical} have considered  $\alpha -\mu $ channel fading to model the small scale  effect in the  THz wireless systems. The  $\alpha -\mu $ is a generalized model that includes other fading models as a special case. It should be noted that the $\alpha-\mu$ model is well studied in the RF context. However, the THz transmission is also impacted by the pointing errors whose effect is inevitable even with highly directional signal beams. The authors in \cite{Boulogeorgos_Analytical} have considered the statistical model of the pointing errors developed for the FSO system to investigate the THz system \cite{Farid2007}. It requires novel approaches to analyze the combined effect of pointing errors and the small scaling fading on the performance of the THz wireless system.

The authors in  \cite{Boulogeorgos_Analytical} have derived distribution functions of the combined channel to analyze the ergodic capacity and outage probability for a single-link  THz system.  However, the cumulative distribution function (CDF) is valid only for integer values of the fading parameter $\mu$.  Recently, a mixed THz-RF link with decode and forward (DF) relaying has been considered in  \cite{Boulogeorgos_Error}. They have derived the average bit-error-rate (BER) and the outage probability of the mixed link using the CDF of \cite{Boulogeorgos_Analytical}. The outage probability of a THz-THz relaying scheme for a multiple antenna system is presented in \cite{Boulogeorgos_20020_THz_Relaying}. The use of relaying for nanoscale communications at THz frequencies has also been presented in \cite{Rong_2017,Abbasi_2017}. It should be noted that there has been extensive research on relaying for RF systems and  RF mixed with the FSO systems \cite{Nosratinia2004,Bjornson_2013,Safari2008}.  However, there is no analysis available in the open literature for the ergodic rate and the average SNR performance for the THz-RF relaying.  Performance bounds on the ergodic rate and average SNR are  desirable for real-time tuning of system parameters for  efficient deployment of THz-RF systems.  

In this paper, we analyze the performance of a THz-RF link with DF relaying for data transmission between the central processing unit of a core network and a user through an AP in a  wireless network. First, we derive a closed-form expression of the CDF of the SNR for THz link using the $\alpha-\mu$ fading channel and statistical models of pointing errors. The derived CDF is also valid for non-integer values of $\mu$ for a generalized performance analysis over fading channels. Next, we derive analytical expressions of the end-to-end SNR and lower bound on ergodic capacity of the THz-RF link in terms of system parameters. We validate the analytical results using numerical and Monte Carlo simulations and show that the THz-RF link performs exceedingly well compared with the direct transmissions.
\section{System Model}
	\label{sec:system_model}	
	The system model consists of a dual-hop relaying, where a DF relay (i.e., an access point (AP)) is placed between the source and destination. The link between the source (a central processing unit of core network) and AP is established by THz transmission, and the AP communicates  a user with  RF link.    In the first hop,  the received signal at the relay is presented as 	
	\begin{equation}
	y_1 = h_{1}h_{pf}s + w_1
	\end{equation}		
	where $s$ is the transmitted   symbol  in the THz band,  $h_{1}$ is channel coefficient,  $h_{pf}$ is the combined  effect of short-term  fading and antenna misalignment, and $w_1$ is the additive noise with variance $\sigma_{w1}^2$. The path gain $h_l$ is dependent on antenna gains, frequency, and  molecular absorption coefficient, as given in \cite{Boulogeorgos_Analytical}. Using the generalized $\alpha-\mu$ fading \cite{Yacoub_alpha_mu} and statistical model of pointing errors   \cite{Farid2007}, the probability distribution function (PDF) of $|h_{pf}|$ is given as \cite{Boulogeorgos_Analytical}:
		\begin{eqnarray} \label{eqn:pdf_hfp_thz}
	f_{|h_{fp}|}(x) = A_1  x^{\phi-1} \times \Gamma (B_1, C_1x^\alpha )
	\end{eqnarray}	
	where $ A_1 = \phi S_0^{-\phi} \frac{\mu^{\frac{\phi}{\alpha}}}{\Omega^\alpha \Gamma (\mu)} $, $ B_1 = \frac{\alpha \mu - \phi}{\alpha} $, and $C_1 =  \frac{\mu}{\Omega^\alpha} S_0^{-\alpha}$.

In the second hop,  assuming that the signal received through the direct link is negligible,  the  signal $y_2$ at the destination is given by 	
	\begin{equation}
	y_2 = h_{2}h_f\hat{s} + w_2
	\end{equation}	
where $\hat{s}$ is decoded symbol at the relay, $ h_2 $ is the channel coefficient,  $w_2$ is the additive noise with variance $\sigma_{w2}^2$, and  $h_f$ is the  short-term  fading with $\alpha-\mu$ distribution for  the RF link as	 
	 \begin{equation} \label{eqn:pdf_hf_rf}
	 f_{|h_f|}(x) = \frac{A_2 x^{\alpha\mu-1}}{\Gamma (\mu)} \exp (-B_2 x^\alpha)
	 \end{equation}	 
	 where $A_2 = \frac{ \mu^\mu}{\Omega^{\alpha\mu}}$ and $B_2 = \frac{\mu}{\Omega^{\alpha}}$.	
 
	 We denote  SNR   of  the THz link as 	$ \gamma_{1}^0|h_{fp}|^2$ and  the RF
as   $\gamma_{2}^0|h_{f}|^2$
	 where $\gamma_{1}^0= \frac{P_1 |h_{1}|^2}{\sigma_{w_1}^2}$ and  $\gamma_{2}^0= \frac{P_2 |h_{2}|^2}{\sigma_{w_2}^2}$ are the SNR terms  without fading for the THz and RF links, respectively. We denote $B_2' = B_2 \sqrt{\frac{\gamma_1^0}{\gamma_2^0}}$ and $C_1' = C_1 \sqrt{\frac{\gamma_2^0}{\gamma_1^0}}$.
\section{Performance Analysis}
First, we derive a closed-form expression on the CDF of THz link over $\alpha-\mu$ fading channel with pointing error. In contrast to \cite{Boulogeorgos_Error}, the derived CDF is also valid for non-integer values of $\mu$. Next, we derive analytical expressions of average SNR and  ergodic rate for the relay-assisted system. 

Using  \eqref{eqn:pdf_hfp_thz}, we can represent the PDF of THz link in terms  of SNR \cite{Boulogeorgos_Analytical}:
\begin{eqnarray}
f_1(\gamma) = \frac{A_1}{2\sqrt{\gamma \gamma_1^0}} \Big(\sqrt{{\gamma}/{\gamma_1^0}}\Big)^{\phi-1} \Gamma \Big(B_1, C_1 \Big(\sqrt{{\gamma}/{\gamma_1^0}}\Big)^\alpha \Big)
\label{eqn:pdf_thz}
\end{eqnarray}
Similarly, we can express \eqref{eqn:pdf_hf_rf} in terms of  SNR for the RF link as
\begin{eqnarray}
f_2(\gamma) \hspace{-1mm}= \frac{A_2 \alpha}{2 \Gamma(\mu)\sqrt{\gamma\gamma_2^0}} \Big(\sqrt{{\gamma}/{\gamma_2^0}}\Big)^{(\alpha\mu-1)} \large{e}^{-B_2 \Big(\sqrt{{\gamma}/{\gamma_2^0}}\Big)^{\alpha}}
\label{eqn:pdf_rf}
\end{eqnarray}
Finally, the CDF of the SNR for the RF link:
\begin{eqnarray}
F_2(\gamma) =  1-\bigg(\frac{\Gamma\big(\mu, B_2 \big(\sqrt{{\gamma}/{\gamma_2^0}}\big)^{\alpha}\big)}{\Gamma (\mu)}\bigg)
\label{eqn:cdf_rf}  
\end{eqnarray}

In the following, we derive a closed-form expression on the CDF of the THz Link
	\begin{my_lemma} If $\phi$ and $S_0$ be the parameters of  pointing errors, and  $\alpha$ and $\mu$ are the fading parameters, then the CDF of the THz link  is given by 
	\begin{eqnarray}
	F_1 (\gamma)=  \frac{A_1  C_1^{-\frac{\phi}{\alpha}}}{\phi} \Big[ \Gamma(\mu) + \Big(C_1\Big(\sqrt{{\gamma}/{\gamma_1^0}}\Big)^{\alpha}\Big)^\frac{\phi}{\alpha}\nonumber \\  \times \Gamma\Big(B_1,C_1\Big(\sqrt{{\gamma}/{\gamma_1^0}}\Big)^{\alpha}\Big) - \Gamma\Big(\mu,C_1\Big(\sqrt{{\gamma}/{\gamma_1^0}}\Big)^{\alpha}\Big) \Big]
	\label{eqn:cdf_thz}
	\end{eqnarray}		
\end{my_lemma}
\begin{IEEEproof} Using  \eqref{eqn:pdf_thz} and substituting $\Big(\sqrt{\frac{\gamma}{\gamma_1^0}}\Big)^\alpha = t$ ,  the CDF of the SNR for THz link is  given by
 \begin{equation}
 F_1 (\gamma) = \frac{A_1}{\alpha} \int_{0}^{\big(\sqrt{{\gamma}/{\gamma_1^0}}\big)^\alpha} t^{(\frac{\phi}{\alpha}-1)} \Gamma (B_1,C_1 t) dt 
 \label{eq:simp}
\end{equation}
To solve the above integral, we use  the following identity:	
\begin{eqnarray}
\int x^{b-1} \Gamma(s, x) \mathrm{d} x= -\frac{1}{b}\big(x^b\Gamma(s,x)+\Gamma(s+b,x)\big)
\label{eq:gamma_inc_identity}
\end{eqnarray} 
Using the limits of  \eqref{eq:simp} in \eqref{eq:gamma_inc_identity}, we get \eqref{eqn:cdf_thz}.
\end{IEEEproof}
 
 The end-to-end SNR for the combined THz-RF DF system is $ \gamma = \mbox{min}(\gamma_1,\gamma_2)$, and thus  $F(\gamma) = F_1(\gamma)+F_2(\gamma)-F_1(\gamma)F_2(\gamma)$ \cite{papoulis_2002}. Using $\gamma=\gamma_{\rm th}$ in the CDF, we can get an expression of the outage probability.
 \subsection{Average SNR}
Using the PDF $f(\gamma)= \frac{d}{d \gamma} F(\gamma)$ \cite{papoulis_2002}, the  average SNR for the relay assisted system is
 \begin{equation} \label{eq:total_pdf}
 {\gamma} = \int_{0}^{\infty}  \gamma [f_1(\gamma)+f_2(\gamma)-f_1(\gamma)F_2(\gamma)-f_2(\gamma)F_1(\gamma)]d \gamma
 \end{equation} 
\begin{figure*}	
	\begin{flalign} 
	\eta_1&= \frac{A_1 C_1^{-\frac{\phi}{\alpha}} \Gamma(\mu)\big(\hspace{-0.5mm}-\hspace{-0.5mm}2(\alpha+\phi \rm log(C_1)) + \alpha\phi \rm log(\gamma_1^0) + 2\phi\psi(0,\mu) \big)}{\alpha \phi^2 \rm log(2)},~~~~~~~ 
	\eta_2= \frac{-2  \log(B_2)+ \alpha \rm log(\gamma_2^0) + 2\psi(0,\mu)}{\alpha \rm log(2)} 
	\label{eq:eta_1_and_2}\\ \eta_{12}  &=\eta_1+\sum_{k=0}^{\mu-1} \frac{2A_1 B_2'^{-\frac{\phi}{\alpha}}}{\alpha^2 \rm log(2) k!}  G_{0,0:2:0}^{0,1:4:2}\bigg[\bigg(\begin{matrix}	1-\frac{\phi}{\alpha}-K \\ - \end{matrix} \bigg |\begin{matrix} 1 \\ B_1, 0 \end{matrix} \bigg |\begin{matrix} 1,1 \\ 1, 0 \end{matrix} \bigg| \frac{C_1}{B_2'},\frac{(\gamma_1^0)^{\frac{\alpha}{2}}}{B_2'}\bigg) - \bigg( \begin{matrix}	1-\frac{\phi}{\alpha}-K \\ - \end{matrix} \bigg |\begin{matrix} 1 \\ B_1, 0 \end{matrix} \bigg |\begin{matrix} 1,1 \\ 1, 0 \end{matrix} \bigg| \frac{C_1}{B_2'},\bigg(\frac{-(\gamma_1^0)^{\frac{\alpha}{2}}}{B_2'}\bigg)^{-1}\bigg) \bigg] \label{eq:eta_12}\\
	\eta_{21} &=  \frac{\big(A_1 C_1^\frac{\phi}{\alpha} \Gamma(\mu)\big) }{\phi}\eta_2  + \frac{2A_1A_2 C_1^{-\frac{\phi}{\alpha}}  C_1'B_2^{-(\mu+\frac{\phi}{\alpha})}}{\Gamma(\mu)\phi{ \log(2)}\alpha}   G_{0,0:2:0}^{0,1:4:2} \bigg[\bigg(\begin{matrix}	1-\mu-\frac{\phi}{\alpha} \\- \end{matrix} \bigg |\begin{matrix} 1 \\ B_1, 0 \end{matrix} \bigg |\begin{matrix} 1,1 \\ 1, 0 \end{matrix} \bigg| \frac{C_1'}{B_2},\frac{(\gamma_2^0)^{\frac{\alpha}{2}}}{B_2}\bigg) \nonumber \\ &- \bigg( \begin{matrix}	1-\mu-\frac{\phi}{\alpha} \\ - \end{matrix} \bigg |\begin{matrix} 1 \\ B_1, 0 \end{matrix} \bigg |\begin{matrix} 1,1 \\ 1, 0 \end{matrix} \bigg| \frac{C_1'}{B_2},\frac{-(\gamma_2^0)^{\frac{\alpha}{2}}}{B_2}\bigg) \bigg]   -\sum_{k=0}^{\mu-1} \frac{2A_1A_2  C_1^{-\frac{\phi}{\alpha}}C_1'^k (B_2+C_1')^{\mu+k}}{\alpha \phi{\rm log(2)}k!}  \nonumber \\& \times G_{3,2}^{1,3} \bigg[\bigg(\begin{matrix} 1,1,1-\mu \\ 1,0 \end{matrix} \bigg|\frac{(\gamma_2^0)^{\frac{\alpha}{2}}}{B_2+C_1'} \bigg) - 	\ \bigg(\begin{matrix} 1,1,1-\mu \\ 1,0 \end{matrix} \bigg| \bigg(\frac{(\gamma_2^0)^{\frac{\alpha}{2}}}{B_2+C_1'} \bigg)^{-1} \bigg) \bigg] \label{eq:eta_21}
	\end{flalign}
	\hrulefill
\end{figure*}
 \begin{my_theorem}
 If $\phi$ and $S_0$ be the parameters of  pointing errors, and  $\alpha$ and $\mu$ are the fading parameters, then the average SNR of the relay assisted THz-RF link is given as
 \begin{eqnarray}  \label{eqn:total_snr}
 	{\gamma} =\gamma_{1}+\gamma_2-\gamma_{12}-\gamma_{21}
 \end{eqnarray} 		
 	where 
 \begin{equation} 
 	\gamma_1 =  \frac{A_1C_1^{-(\frac{\phi+2}{\alpha})} \gamma_1^0 \Gamma(\frac{\alpha B_1+\phi+2}{\alpha})}{\phi+2},   \gamma_2 = \frac{B_2^{-\frac{2}{\alpha}} \gamma_2^0 \Gamma(\frac{2}{\alpha}+\mu)}{\Gamma(\mu)}
 \label{eq:gamma_1_and_2}
 \end{equation}
 \begin{eqnarray} \label{eq:gamma_12}
	  & 	\gamma_{12}  =  \gamma_1 - \frac{1}{(\phi+2)\Gamma(\mu)}\bigg[A_1 B_2'^{-\frac{\phi+2}{\alpha} }\gamma_{1}^0 \bigg(\Gamma(B_1) \Gamma\left(\frac{\alpha\mu+\phi+2}{\alpha}\right) \nonumber \\& +B_2'^{-B_1} C_1^{B_1} \Gamma\left(\frac{\alpha(B_1+\mu)+\phi+2}{\alpha}\right) \nonumber \\& \times \bigg(\frac{\alpha ~ _2\tilde{F}_1\big[\frac{\alpha B_1+\phi+2}{\alpha}, \frac{\alpha(B_1+\mu)+\phi+2}{\alpha}, \frac{\alpha B_1+\alpha+\phi+2}{\alpha}, -\frac{C_1}{B_2'} \big]}{\alpha B_1+\phi+2} \nonumber \\&  - \Gamma(B_1)~ _2\tilde{F}_1 \bigg[B_1, \frac{\alpha(B_1+\mu)+\phi+2}{\alpha}, 1+B_1, -\frac{C_1}{B_2'} \bigg] \bigg) \bigg) \bigg]
 \end{eqnarray}
 \begin{eqnarray}  \label{eq:gamma_21}
 	&\gamma_{21} =  \frac {A_1 C_1^{\frac{-\phi}{\alpha}} \Gamma(\mu)  \gamma_{2}}{\phi} + \frac{1}{\phi\Gamma(\mu)}\bigg[\alpha A_1 A_2 C_1^{-\frac{\phi}{\alpha}} C_1'^{-\frac{\alpha\mu+\alpha+2}{\alpha}} \gamma_2^0 \nonumber \\& \times \bigg(-\frac{\Gamma(1+ \frac{2}{\alpha}+2\mu) ~ _2\tilde{F}_1\big[1+ \frac{2}{\alpha}+\mu, 1+ \frac{2}{\alpha}+2\mu, 2+ \frac{2}{\alpha}+\mu, -\frac{B_2}{C_1'} \big]}{\alpha \mu +\alpha	+2} \nonumber \\ & + [\alpha \mu +\alpha + \phi +2]^{-1}\Gamma(\frac{\alpha(\mu+B_1+1)+\phi+2}{\alpha})  _2\tilde{F}_1 \nonumber\\&  \times\big[\frac{\alpha\mu+\alpha+\phi+2}{\alpha}, \frac{\alpha(\mu+B_1+1)+\phi+2}{\alpha}, \frac{\alpha(\mu+2)+\phi+2}{\alpha}, -\frac{B_2}{C_1'} \big] \bigg) \bigg] 	
 \end{eqnarray} 				
 \end{my_theorem}
 \begin{IEEEproof}
 	The proof is presented in Appendix A.
 \end{IEEEproof}
\subsection{Ergodic Capacity} 
Similar to the average SNR, and using the identity $\log_2(1+\gamma)\geq \log_2(\gamma)$, a lower bound on the ergodic capacity is defined:

 \begin{equation}
 \eta=\int_{0}^{\infty} \log_2(\gamma) [f_1(\gamma)+f_2(\gamma)-f_1(\gamma)F_2(\gamma)-f_2(\gamma)F_1(\gamma)]d \gamma
 \end{equation}	
\begin{my_theorem}  If $\phi$ and $S_0$ be the parameters of  pointing errors, and  $\alpha$ and $\mu$ are the fading parameters, then the average ergodic capacity of the relay assisted THz-RF link is given as 
	\begin{eqnarray}
	\eta =\eta_1+\eta_2-\eta_{12}-\eta_{21}
	\end{eqnarray}	
where $ \eta_1$ and $\eta_2,\eta_{12}$, and $\eta_{21} $ are given in \eqref{eq:eta_1_and_2}, \eqref{eq:eta_12}, and \eqref{eq:eta_21}, respectively.

\end{my_theorem}	
\begin{IEEEproof}The proof is presented in Appendix B.
\end{IEEEproof}

 \begin{figure*}[tp]
	\begin{center}
		\subfigure[Outage probability for different values of $\gamma_{th}$  ]{\includegraphics[scale=0.24]{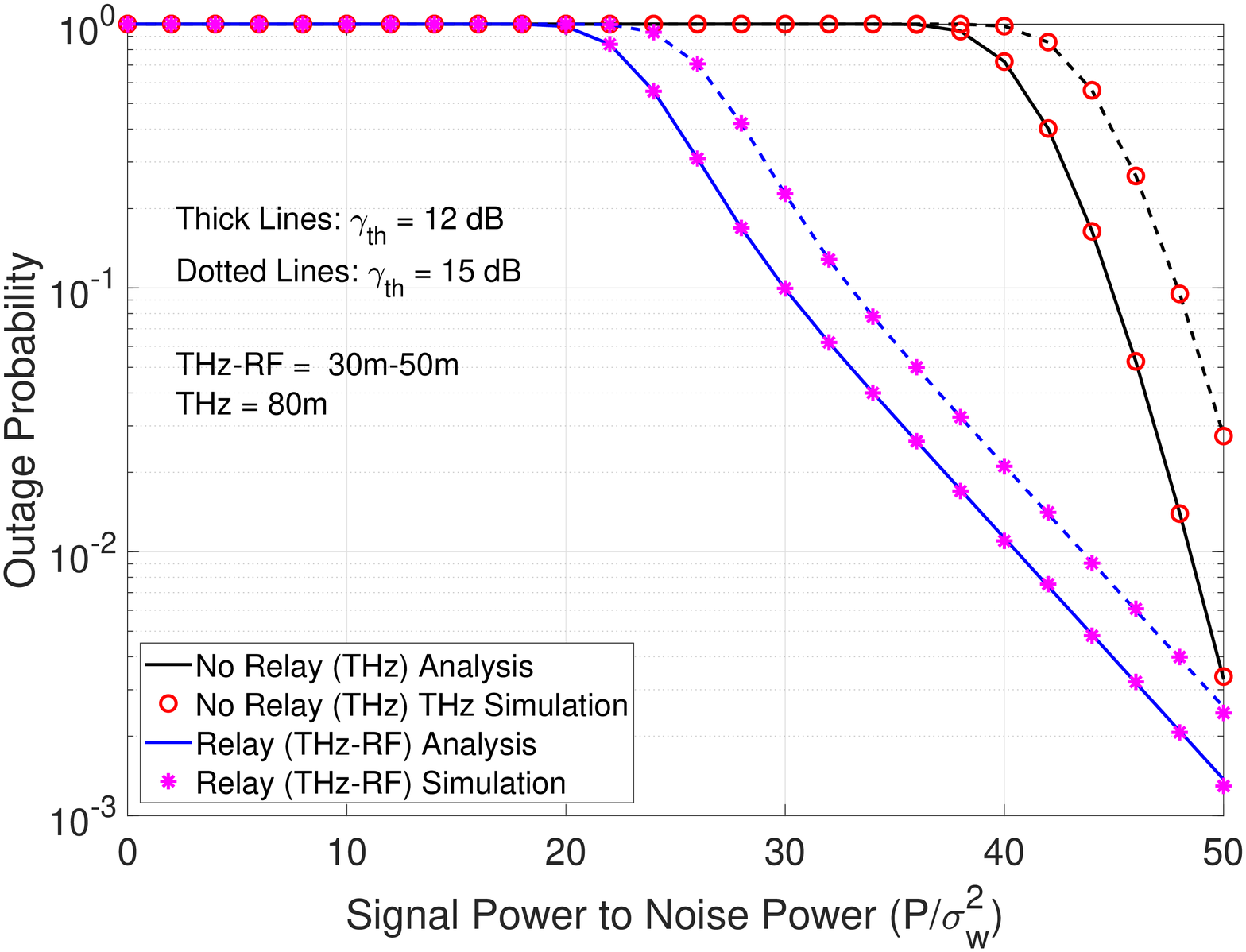}}
		\subfigure[Average SNR.]{\includegraphics[scale=0.24]{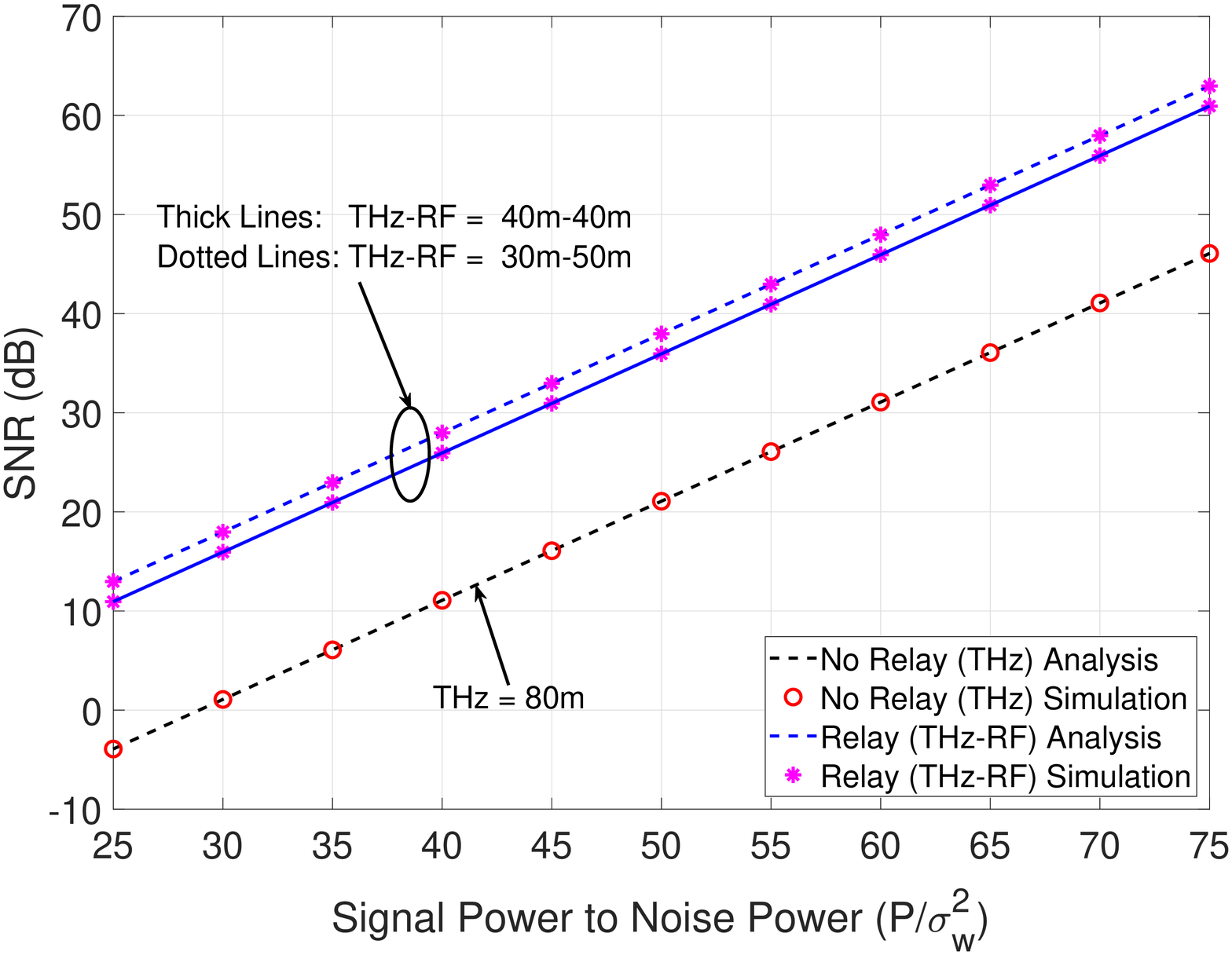}}
		\caption{Performance comparison of relay-assisted THz-RF wireless link with the direct THz transmission.}
		\label{outage}
	\end{center}
\end{figure*}

\section{Simulation and Numerical Results}
 In this section, we use numerical analysis and Monte Carlo simulations (averaged over $10^6$ channel realizations) to demonstrate the performance of the THz-RF relay assisted system. We adopt similar  simulation parameters as given in \cite{Boulogeorgos_Error,Boulogeorgos_Analytical}. We consider the THz link with a distance of  $40$ \mbox{m}, carrier frequency  $275$ \mbox{GHz},  and  antenna gain of  $55$ \mbox{dBi}. To compute the path loss for the THz link, we consider the relative humidity, atmospheric pressure,  and temperature as  $ 50\% $,  \mbox{101325} Pa, and 296\textdegree K, respectively.   The RF link distance is taken upto 50 m, carrier frequency 2 GHz, and antenna gain are 36 dBi. We compute the path loss of the RF link  using the 3GPP model $h_2 = 32.4+17.3\log_{10}(d_2)+20\log_{10} (10^{-9}f_2)$,  where $d_2$ is the distance and $f_2$ is the carrier frequency of RF link.  To simulate the $\alpha-\mu$  fading channel, we  take $ \alpha = 2 $ and $ \mu=4 $. Using measurement data in
\cite{Farid2007}, the  pointing error parameters are $\phi=8.5448$ and $S_0=0.1172$. We also compute achievable data rate of the relay-assited system by  considering power spectral density (PSD) of  noise as   $3.8\times10^{-17} \rm{W/Hz}$ over a channel bandwidth of $10$ \mbox{GHz} \cite{Sen_2020_Teranova}.

\begin{figure}	
 	{\includegraphics[scale=0.24]{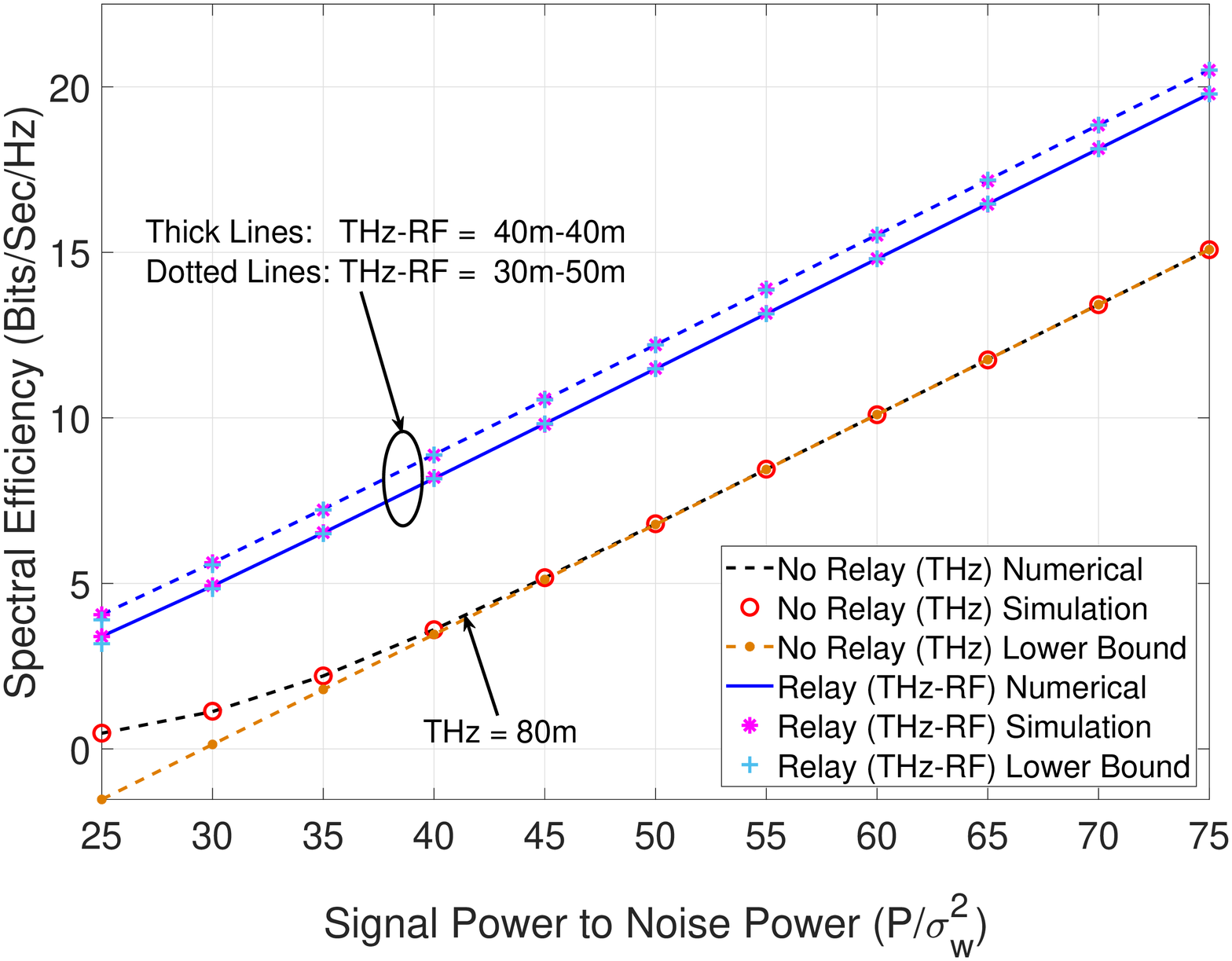}}		
 	\caption{Spectral efficiency of relay-assisted THz-RF wireless link, and its comparison with the direct THz transmission.}
 	\label{rate}
 \end{figure}

In Fig.~\ref{outage}(a), we demonstrate the outage performance of the THz-RF link for different SNR thresholds $ \gamma_{th}=12 $ \mbox{dB} and $ \gamma_{th}=15 $ \mbox{dB}. It can be seen that the outage performance of the relaying is significantly better than direct transmission. The figure shows that the outage probability of the relay-assisted scheme is  $100$ times better than the direct THz transmission at a $40$\mbox{dB} of signal power to noise power ratio, which corresponds to $6$\mbox{dBm} of signal power. We also compare the average SNR performance of the THz-RF link with the direct THz transmission, as shown in Fig.~\ref{outage}(b). The figure shows that the relay assisted scheme achieves a significantly higher average SNR performance (more than $ 15 $ \mbox{dB}) over the entire range of $P/\sigma_w^2$. Finally, to demonstrate the applicability of the THz-RF link for the back-haul, in Fig.~\ref{rate}, we show the spectral efficiency performance of the THz-RF link. Figure shows a notable gain of $5$ \mbox{bits/sec/Hz} in the spectral efficiency of the relay assisted system.

Although the average SNR and ergodic capacity of the relay-assisted system show a consistent improvement than the direct transmission over a wide range of $P/\sigma_w^2$, the outage performance shows a decrease in the performance.  The dependence of outage probability on the $P/\sigma_w^2$ ratio and decrease in the performance of relay assisted system compared to the direct transmission is attributed to the independent fading in the RF and THz channels, and the use of DF relaying $\gamma= \min(\gamma_{1},\gamma_{2})$. Since the channels are highly random, the instantaneous SNR $\gamma$ can be worse than the direct THz channel. We have also demonstrated that the relaying with unequal link distances of the source to relay and relay to destination performs slightly better than the relaying with equal distances. The plots also show that our derived analytical results and bounds are in good agreement with the numerical and simulation results.
\section{Conclusion}
We have analyzed the performance of the THz-RF relay link for back-haul applications in the next generation of cellular networks. By deriving a closed-form expression of the CDF for THz link using the $\alpha-\mu$ fading and statistical models of pointing errors, we have derived a closed-form expression of the average SNR and a lower bound on the ergodic capacity for the THz-RF link. Simulation and numerical results show a significant gain (more than $25\%$ increase) in the performance of average SNR and spectral efficiency of the relay-assisted system compared with the direct transmissions. The THz-RF link can achieve higher data rates, which can be sufficient for data transmission between users and the central processing unit in a cell-free wireless network.
\section*{Appendix A: Proof of theorem 1}
Using (\ref{eqn:pdf_thz}), we define $\gamma_1= \int_{0}^{\infty} \gamma f_1(\gamma) d\gamma$, and substituting $\Big(\sqrt{\frac{\gamma}{\gamma_1^0}}\Big)^\alpha = t$, we get the average SNR of the THz link
\begin{eqnarray} 
\label{eq:gamma_1_int}
\gamma_{1} = \frac{A_1 \gamma_1^0}{\alpha} \int_{0}^{\infty} t^{(\frac{\phi-\alpha+2}{\alpha})} \times \Gamma (B_1,C_1t) dt 
\end{eqnarray}
Using the identity \cite{DLMF} in  \eqref{eq:gamma_1_int}, we solve the integral to get $\gamma_{1}$ of \eqref{eq:gamma_1_and_2}. Similarly,  average SNR of the RF link is $\gamma_2= \int_{0}^{\infty} \gamma f_2(\gamma) d\gamma$,  which can be solved easily  to  get \eqref{eq:gamma_1_and_2}.
 Using \eqref{eqn:pdf_thz} and \eqref{eqn:cdf_rf} in  $\gamma_{12}= \int_{0}^{\infty} \gamma f_1(\gamma) \gamma F_1(\gamma) d\gamma$, and substituting $\Big(\sqrt{\frac{\gamma}{\gamma_1^0}}\Big)^\alpha = t$, we get
\begin{flalign} \label{eq:gamma_12_int}
\gamma_{12} &= \frac{A_1 \gamma_1^0}{\alpha} \bigg[\int_{0}^{\infty} t^{(\frac{\phi-\alpha+2}{\alpha})} \Gamma (B_1,C_1 t)dt \nonumber \\ &- \int_{0}^{\infty} t^{(\frac{\phi-\alpha+2}{\alpha})} \Gamma (B_1,C_1 t) \left(\frac{\Gamma\left(\mu, B_2' t\right)}{\Gamma (\mu)}\right)dt\bigg] 
\end{flalign}
where $B_2' = B_2 \left(\sqrt{\frac{\gamma_1^0}{\gamma_2^0}}\right)$. The first integral in \eqref{eq:gamma_12_int} is same as \eqref{eq:gamma_1_int}.  To solve the second integral,  we apply the integration by parts tasking  $\Gamma\left(\mu, B_2' t\right)$ as the first and $t^{(\frac{\phi-\alpha+2}{\alpha})} \Gamma (B_1,C_1 t)$ as the second term, and use the identity  [\cite{Gradshteyn},eq.(6.455/1)] to get \eqref{eq:gamma_12}. Similarly, using (\ref{eqn:pdf_rf}) and (\ref{eqn:cdf_thz}) in  $\gamma_{21}= \int_{0}^{\infty} \gamma f_2(\gamma) \gamma F_1(\gamma) d\gamma$, and substituting $\Big(\sqrt{\frac{\gamma}{\gamma_2^0}}\Big)^\alpha = t$, we get
\begin{flalign} \label{eq:gamma_21_int}
\gamma_{21}&= \frac{A_1A_2C_1^{-\frac{\phi}{\alpha}}\gamma_2^0}{\Gamma(\mu)\phi} \bigg[ \int_{0}^{\infty} t^{(\mu-1+\frac{2}{\alpha})} \exp (-B_2 t) \Gamma(\mu) dt \nonumber \\ & +\int_{0}^{\infty} t^{(\mu-1+\frac{2}{\alpha})} \exp (-B_2 t) C_1' t^{\frac{\phi}{\alpha}}  \Gamma(B_1,C_1' t) dt \nonumber \\ &-\int_{0}^{\infty} t^{(\mu-1+\frac{2}{\alpha})} \exp (-B_2 t) \Gamma(\mu,C_1't) dt \bigg]
\end{flalign}
where $C_1' = C_1 \left(\sqrt{\frac{\gamma_2^0}{\gamma_1^0}}\right)$. The first integral in \eqref{eq:gamma_21_int} is similar to $\gamma_{2}$ and can be derived likewise. For the second and third integrals, we use the identity 
 [\cite{Gradshteyn},eq.(6.455/1)] to get \eqref{eq:gamma_21}. 
\section*{Appendix B: Proof of theorem 2}
Using (\ref{eqn:pdf_thz}), we define $\eta_1 = \int_{0}^{\infty} {\log}_2(\gamma) f_1(\gamma)d\gamma$, and substituting $\Big(\sqrt{\frac{\gamma}{\gamma_1^{(0)}}}\Big)^\alpha = t$, we get average capacity of THz link:	
\begin{equation} \label{eq:eta_1_int}
\eta_1 = \frac{A_1}{\alpha  \log(2)}\int_{0}^{\infty} {\rm log} (\gamma_1^0 t^{\frac{2}{\alpha}}) t^{(\frac{\phi}{\alpha}-1)}  \Gamma (B_1,C_1 t) dt 
\end{equation}	
To find a closed-form expression,  we use integration by parts taking $\Gamma (B_1,C_1 t)$ as the first and $ {\rm log} (\gamma_1^0 t^{\frac{2}{\alpha}}) t^{(\frac{\phi}{\alpha}-1)}$ as the second term,  and apply the identity [\cite{Gradshteyn}(eq.4.352/1)] to get $\eta_1$ of \eqref{eq:eta_1_and_2}. Similarly, substituting $\Big(\sqrt{\frac{\gamma}{\gamma_2^0}}\Big)^\alpha=t$ in average capacity of the RF link $\eta_2= \int_{0}^{\infty} \rm{log}_2(\gamma) f_2(\gamma)d\gamma$, we get
\begin{equation} \label{eq:eta_2_int}
\eta_2 = \frac{A_2}{\Gamma(\mu) \rm log(2)}\int_{0}^{\infty} { \log} (\gamma_2^0 t^{\frac{2}{\alpha}}) t^{(\mu-1)} \exp (-B_2 t) dt
\end{equation}	
We use the identity [\cite{Gradshteyn}(eq.4.352/1)] in \eqref{eq:eta_2_int} to get $\eta_2$ of \eqref{eq:eta_1_and_2}. Defining $\eta_{12} = \int_{0}^{\infty} {{\log}}_2(\gamma) f_1(\gamma) F_2(\gamma)d\gamma$, using \eqref{eqn:pdf_thz} and \eqref{eqn:cdf_rf}, and substituting $\bigg(\sqrt{\frac{\gamma}{\gamma_1^0}}\bigg)^\alpha = t$ we get
\begin{eqnarray} \label{eq:eta_12_int}
\eta_{12} = \frac{A_1}{\alpha \rm log(2)} \bigg[\int_{0}^{\infty} {\rm log} (\gamma_1^0t^{\frac{2}{\alpha}}) t^{(\frac{\phi}{\alpha}-1)} \Gamma (B_1,C_1 t) dt \nonumber \\ - \int_{0}^{\infty} {\rm log} (\gamma_1^0 t^{\frac{2}{\alpha}}) t^{(\frac{\phi}{\alpha}-1)} \Gamma (B_1,C_1 t) \left(\frac{\Gamma\left(\mu, B_2' t\right)}{\Gamma (\mu)}\right) dt \bigg]  
\end{eqnarray}
The first integral in \eqref{eq:eta_12_int} is same as \eqref{eq:eta_1_int}. To solve the second integral we use the series expansion of Gamma function $\Gamma(\mu,B_2't) = (\mu-1)! \exp(-B_2't) \sum_{k=0}^{\mu-1}\frac{(B_2't)^k}{k!}$. Further, we use the  Meijer's G representation of ${\rm log} (\gamma_1^0 t^{\frac{2}{\alpha}}) $, $\Gamma (B_1,C_1 t)$ and $\Gamma\left(\mu, B_2' t\right) $ to get the second integral as
\begin{flalign}
I_2 &= \sum_{k=0}^{\mu-1} \frac{2A_1(\mu-1)! B_2'^k}{\alpha^2 \rm log(2) \Gamma(\mu) k!} \int_{0}^{\infty} t^{(\frac{\phi}{\alpha}+k-1)} G_{0,1}^{1,0} \Big(\begin{matrix} - \\ 0 \end{matrix} \Big|B_2' t \Big)  \nonumber \\ &\times G_{1,2}^{2,0} \Big(\begin{matrix} 1 \\ B_1,0 \end{matrix} \Big|C_1 t \Big)  \Big[G_{2,2}^{1,2}\Big(\begin{matrix} 1,1 \\ 1,0 \end{matrix} \Big|(\gamma_1^0)^{\frac{\alpha}{2}} t \Big) \nonumber \\ &- G_{2,2}^{1,2} \Big( \begin{matrix} 1,1 \\ 1,0 \end{matrix} \Big|\left((\gamma_1^0)^{\frac{\alpha}{2}} t\right)^{-1} \Big) \Big] dt
\end{flalign}
Finally, we apply the identity of definite integration of the product of three Meijer's G function \cite{Mathematica_three} to get \eqref{eq:eta_12}. Similarly, $\eta_{21} = \int_{0}^{\infty} {\log}_2(\gamma) f_2(\gamma) F_1(\gamma)d\gamma$ can be expressed as
\begin{flalign}  \label{eq:eta_21_int}
\eta_{21} &= \frac{A_1  C_1^{-\frac{\phi}{\alpha}} A_2}{\Gamma(\mu)\phi {\rm log(2)}} \bigg[\int_{0}^{\infty} {\rm log} ( \gamma_2^0 t^{\frac{2}{\alpha}}) t^{(\mu-1)} \exp (-B_2 t)  \Gamma(\mu)dt  \nonumber \\ &+ \int_{0}^{\infty} {\rm log} ( \gamma_2^0 t^{\frac{2}{\alpha}}) t^{(\mu-1)} \exp (-B_2 t) C_1' t^{\frac{\phi}{\alpha}}  \Gamma(B_1,C_1't)dt \nonumber \\ &- \int_{0}^{\infty} {\rm log} ( \gamma_2^0 t^{\frac{2}{\alpha}}) t^{(\mu-1)} \exp (-B_2 t) \Gamma(\mu,C_1't)dt \bigg] 
\end{flalign}

The first integral in \eqref{eq:eta_21_int}  is similar to $\eta_2$ of \eqref{eq:eta_2_int}. To solve the second integral, we apply the identity of definite integration of the product of three Meijer's G function \cite{Mathematica_three}. Finally, to solve the third integration, we use the series expansion of Gamma function $\Gamma(\mu,C_1't) = (\mu-1)! \exp(-C_1't) \sum_{k=0}^{\mu-1}\frac{(C_1't)^k}{k!}$ and we apply the identity of definite integration of the product of two Meijer's G function \cite{Mathematica_two}. Combining these three integration, we get \eqref{eq:eta_21}.
\section*{Acknowledgment}
This work is supported in part by the Science and Engineering Research Board (SERB), Government of India under Start-up Research Grant SRG/2019/002345.

\bibliographystyle{ieeetran}
\bibliography{thz_bib_file}

	\end{document}